\documentclass[aps,prl,showpacs,twocolumn,superscriptaddress]{revtex4}

\usepackage{amssymb}
\usepackage{epsfig}
\usepackage{graphicx}
\usepackage{amsmath}
\usepackage{array,color}

\begin{document}
\title{Superradiant Solid  in Cavity QED
Coupled to a Lattice of Rydberg Gas}

\author{Xue-Feng Zhang}
\affiliation{Physics Department and Research Center OPTIMAS,
University of Kaiserslautern, 67663 Kaiserslautern, Germany}
\author{Qing Sun}
\affiliation{Institute of Physics, Chinese Academy of Sciences,
Beijing 100190, China}
\author{Yu-Chuan Wen}
\affiliation{Department of Physics, Capital Normal University,
Beijing 100048, China}
\author{Wu-Ming Liu}
\affiliation{Institute of Physics, Chinese Academy of Sciences,
Beijing 100190, China}
\author{Sebastian Eggert}
\affiliation{Physics Department and Research Center OPTIMAS,
University of Kaiserslautern, 67663 Kaiserslautern, Germany}

\author{An-Chun Ji}
\email{andrewjee@sina.com} \affiliation{Department of Physics,
Capital Normal University, Beijing 100048, China}

\date{{\small \today}}


\begin{abstract}

We study an optical cavity coupled to a lattice of Rydberg atoms,
which can be represented by a  generalized Dicke model.  We show
that the competition between the atomic interaction and atom-light
coupling induces a rich phase diagram. A novel ``superradiant
solid" (SRS) phase is found, where both the superradiance and
crystalline orders coexist. Different from the normal second order
superradiance (SR) transition, here both the Solid-1/2 and SRS to
SR phase transitions are first order. These results are confirmed
by the large scale quantum Monte Carlo simulations.

\end{abstract}

\pacs{42.50.Pq, 03.75.Nt, 05.30.Jp, 67.10.Fj}





\maketitle

{\it Introduction.}-- The study of quantum many-body problems and
quantum phase transitions (QPT) has attracted great interest and
is currently one of the main issues in the condensed matter
community \cite{Sachdev}. In the past decade, the successful
control of the interaction strength and dimensionality of
ultracold quantum gases has made it possible to explore many
interesting physical phenomena \cite{quantumgas}. For example, the
observation of the superfluid-Mott insulator transition
\cite{Greiner}, a Tonks gas \cite{Paredes}, the BEC to BCS
crossover \cite{Chen}, and the Kosterlitz-Thouless phase
transition \cite{Hadzibabic} have all been established in
ultra-cold quantum gases.

More recently ultracold atoms have been combined with cavity QED
\cite{Brennecke,Colombe} to study atom-light coupled many-body
problems, which give rise to new phenomena.  In particular, when
the two-level atom gas is coupled to a cavity \cite{Baumann} the
coherent non-local atom-light interaction supports the famous
superradiance (SR) phase transition in the  Dicke model (DM)
\cite{Dicke,Hepp}. This SR phase is formed by the condensation of
atom-light coupled polaritons \cite{Esthman} and  breaks the U(1)
symmetry. However, the interactions between atoms, which may bring
new phenomena or induce novel QPTs, are not considered in the  DM.

Atomic interactions promise to give interesting new effects and
can be implemented via Rydberg atoms.  The  unique properties of
strong dipole-dipole interactions and quite long lifetimes have
made them a powerful tool for the implementation of coherent
blockade effects and quantum information \cite{Saffman}.
Especially, the successful trapping of Rydberg atoms in a 1D
optical lattice \cite{Anderson} has stimulated the study of
many-body quantum systems such as spin system
\cite{Olmos,Lesanovsky} and dynamical crystallization or melting
of ultracold atoms \cite{Pohl,Weimer2}.

In this Letter, we now analyze the generalized DM by coupling a 1D
lattice of Rydberg atoms with an optical cavity, where the dipole
interaction between two Rydberg atoms competes with the atom-light
coupling. We see that, while the atom-light interaction favors the
SR phase, the atomic interaction tends to form three
incompressible Rydberg solid states with filling numbers 0, 1/2,
and 1, which destroys the polariton formation. Most importantly,
we find a novel state corresponding to a ``superradiant solid"
(SRS) phase, where both the superradiance and crystalline orders
coexist and the corresponding U(1) and translation symmetries are
broken simultaneously. Compared with the supersolid (SS) phase in
an optical lattice \cite{Sengupta,wess,Batrouni}, which breaks the
same symmetries, the SRS is rather unique as it is induced by the
non-local atom-light coupling and the condensation of  polaritons.
Moreover we find that, while the Solid-$(0,1)$ to SR phase
transitions remain second order, both the Solid-1/2 and SRS to SR
phase transitions become first order.

{\it The model.}-- The system consists of a deep 1D lattice with
$N$ sites in  an optical cavity. Each site is occupied by a single
atom with the ground state $|$g$\rangle$ coupled to a high-lying
Rydberg state $|e\rangle$ via a two-photon transition process
\cite{Guerlin}. The Hamiltonian of this system in the
rotating-wave approximation  is given by
\begin{eqnarray}
\hat{H}\!\!=\!\!&&\omega \psi^\dag \psi + \sum_{i=1}^N
\frac{\epsilon}{2} (b_i^{\dag}b_i-a_i^{\dag}a_i)+\!\!
\frac{g}{\sqrt{N}}\sum_{i=1}^N (b_i^{\dag} a_i\psi +
{\rm H.c.})\nonumber\\
&&+C_6 \sum_{\langle i,j\rangle}P_{ee}^{(i)}P_{ee}^{(j)}-\mu
N_{\rm ex},\label{hamiltonian}
\end{eqnarray}
where $\omega$ and $\epsilon$ are the cavity and atom transition
frequencies with the detuning defined by $\Delta=\omega-\epsilon$.
$\psi^{\dag}$ is the single-mode creation operator of the cavity
field, $a_i$ and $b_i$ are the boson operators representing the
lower and upper levels of each atom and satisfy the
single-occupancy constraint $b_i^{\dag}b_i+ a_i^{\dag}a_i=1$ and
$g_{\rm eff}\equiv g/\sqrt N$ is the effective two-photon
coupling. The strong dipole-dipole interactions $C_6$ between two
Rydberg states is modeled by projectors $P_{ee}^{(i)}=n_i\equiv
b_i^{\dag}b_i$ onto the Rydberg state, where only nearest
interactions are considered \cite{Olmos}. The last term is the
chemical potential for the total number of excitations $N_{\rm
ex}=\psi^\dag \psi+\sum_{i=1}^N b_i^{\dag}b_i$.

The above model  possesses two limiting cases. First for $C_6=0$,
the Hamiltonian (\ref{hamiltonian}) becomes the DM. Here we note
that, because the ground state $|$g$\rangle$ of the atoms is not
directly coupled to the Rydberg state $|e\rangle$, the so called
no-go theorem \cite{Rzazewski} does not apply \cite{Larson}, and
the SR phase can occur. Recently, the SR phase transition has been
observed with a superfluid atomic gas in an optical cavity
\cite{Baumann}. Secondly, when $g$ is zero  this system becomes a
pure lattice of Rydberg atoms. Then by tuning the chemical
potential $\tilde{\mu}\equiv\mu-\omega$, one may derive three
Rydberg solid states: (i) $\tilde{\mu}<-\Delta$, it forms a
Solid-0 phase with all the atoms staying in the ground state; (ii)
$-\Delta<\tilde{\mu}<2C_6-\Delta$, half of the atoms are excited
to the Rydberg states and form a Solid-1/2 phase because of the
nearest neighbor repulsion; (iii) $\tilde{\mu}>2C_6-\Delta$, all
the atoms are excited to the Rydberg states forming a Solid-1
phase.

\begin{figure}[h]
\includegraphics[width=0.45\textwidth]{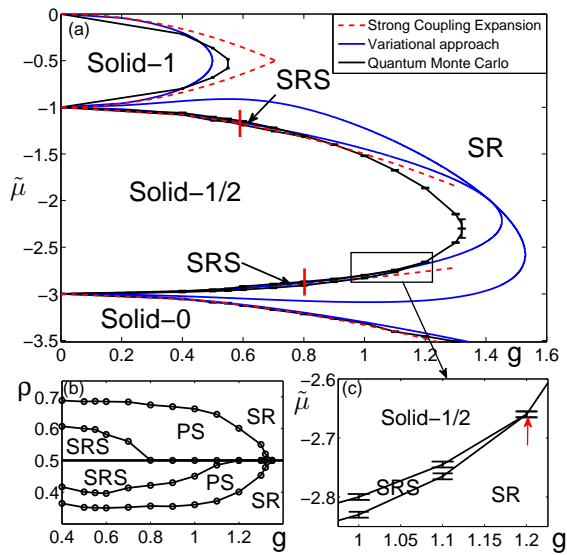}%
\caption{(color online). (a) The phase diagram obtained by
different methods for $C_6=1$ and $\Delta=3$ in the
grand-canonical ensemble. (b) The phase diagram obtained by QMC in
the canonical ensemble. Here, PS denotes the phase separation
between the Solid-1/2, SRS and SR phase transitions. (c) The
enlarged region of the SRS phase, where the red arrow marks the
tricritical point among SRS, SR and Solid-1/2 phases.
\label{mcphase}}
\end{figure}

{\it Analytical Approach.}--A convenient starting point is to
consider the limit $g\ll C_6$. In this case, one may take the
atom-light interaction as a perturbation and use the strong
coupling expansion (SCE) \cite{Freericks}. By comparing the
energies of exciting a particle or hole with the correction from
second order processes in the solid phase, we derive the following
melting critical lines of the incompressible solid lobes
\cite{note1}:
\begin{flushleft}
\begin{tabular}{lll}
Solid-0: & & $\tilde{\mu}_{c1}=-\Delta-g^2/\Delta$, \\
Solid-1/2: & & $\tilde{\mu}_{c2}=-\Delta+g^2/(2\Delta)$, \\
& & $\tilde{\mu}_{c3}=-\Delta+2C_6-g^2/(2\Delta-4C_6)$,\\
Solid-1: & & $\tilde{\mu}_{c4}=-\Delta+g^2/(\Delta-2C_6)+2C_6$, \\
& & $\tilde{\mu}_{c5}=-g^2/(\Delta-2C_6)$,\\
\end{tabular}
\end{flushleft}
which are shown as dashed lines in Fig.~\ref{mcphase}.  The
melting will make the solid compressible, but it needs to be
explored if the order in the Solid-1/2 state is actually destroyed
or may be stable even in the presence of particle or hole
excitations.

\begin{figure}[h]
\includegraphics[width=0.45\textwidth]{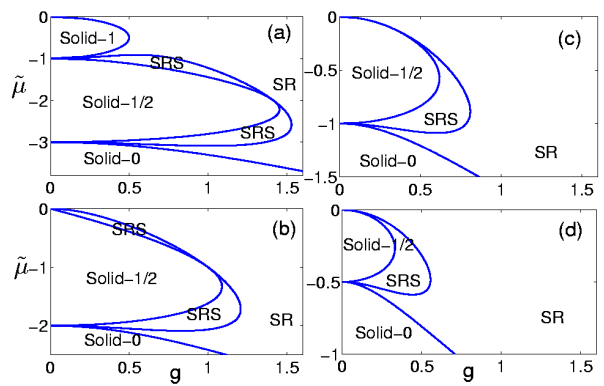}%
\caption{Variational phase diagram for (a) $\Delta=3$, (b)
$\Delta=2$, (c) $\Delta=1$, and (d) $\Delta=0.5$. All parameters
are in units of $C_6$. \label{phase}}
\end{figure}

For this purpose we introduce a variational ground wavefunction
\cite{Esthman} to gain a  more physical insight into the phase
diagram:
\begin{eqnarray}
|\lambda,\theta\rangle=\exp(\frac{\lambda\sqrt{N}\psi^{\dag}}2)
\prod_i[\cos(\frac{\theta_i}2)b_i^{\dag}+\sin(\frac{\theta_i}2)
a_i^{\dag}]|0\rangle,
\end{eqnarray}
where $|0\rangle$ denotes the vacuum state with all atoms in the
ground state, and $\lambda$ and $\theta_i$ are the variational
parameters for the coherent cavity field and atomic fields. To
find the ground state, we calculate the energy density
$\mathcal{E}\equiv4\langle
\lambda,\theta|H|\mathcal{V}|\lambda,\theta\rangle/N$:
\begin{eqnarray}
\nonumber \mathcal{E} &=& 4[ -g\lambda(\sin\theta_A+\sin\theta_B)
+C_6\cos\theta_A\cos\theta_B \\
&&-\tilde{\mu}\lambda^2-(\tilde{\mu}+\Delta-C_6)(\cos\theta_A
+\cos\theta_B)],\label{energy}
\end{eqnarray}
where we have assumed two sublattices $A$ and $B$. From
Eq.~(\ref{energy}) we see that while the  $g$-term tends to
enhance the cavity field $\lambda$ and the condensation of
polaritons, the second $C_6$-term favors the staggered order of
the polaritons. After minimizing $\mathcal{E}$ in respect to the
variational parameters, we derive the phase diagram
Fig.~\ref{phase}. The corresponding variational values are shown
in TABLE \ref{table}, where the solid phases represent the Rydberg
crystals without coherent cavity excitations. The usual
superradiance phase is denoted by SR, where the atoms and light
form polaritons and condense with  a nonzero order parameter
$\langle b_i^{\dag}a_i\rangle=\sin\theta_{\rm SR}/2$ and a
coherent cavity field $\lambda_{\rm SR}$.
\begin{table}[h]
\centering
\begin{tabular}{|c|c|c|c|c|c|}
\hline
Variational values & Solid-0 & SRS & SR & Solid-1/2 & Solid-1 \\
\hline
$\theta_A$ & $\pi$ & $\theta_1$ & $\theta_{\rm SR}$ & 0 & 0\\
\hline
$\theta_B$ & $\pi$ &  $\theta_2$ &  $\theta_{\rm SR}$ &  $\pi$ & 0\\
\hline
$\lambda$ & $0$ &  $\lambda_{\rm SRS}$ &  $\lambda_{\rm SR}$ &  $0$ & 0\\
\hline
\end{tabular}\caption{The variational values for different phases, where
$\theta_1\neq\theta_2$ in SRS phase.} \label{table}
\end{table}
The most interesting finding is that, when $g\sim C_6$ there
exists an intermediate SRS phase between the Solid-1/2 and SR
phases. Different from the SR phase where the polaritons are
excited uniformly, the SRS phase also breaks the translation
symmetry, which shows a characteristic excitation density
$\rho_{A,B}=(\cos\theta_{1,2}+1)/2$ that is not equal on the two
sublattices. Therefore, both the superradiance and crystalline
orders coexist in the SRS phase. In Fig.~\ref{phase}, we also see
that the solid lobes shrink with decreasing $\Delta$.

In bipartite lattices with nearest neighbor repulsion of bosons a
related supersolid (SS) phase is known to exist, but this phase is
only stable for {\it soft-core} contact interaction in a 1D
Bose-Hubbard (BH) model \cite{Batrouni}. It is therefore
surprising to find an SRS phase in our system which is effectively
described by hard-core bosons, since only a single polariton can
be excited in each site. Here we note that, different from the BH
model, there is no direct kinetic hopping term but instead the
non-local cavity field couples with all atoms simultaneously. This
is the unique property that makes the appearance of the SRS rather
nontrivial in this system.

However, the above results are derived from mean-field
calculations, so the SRS phase may not be stable when quantum
effects are taken into account. In particular, when a particle (or
hole) is introduced into the Solid-1/2 phase, the strong quantum
fluctuation of the non-local atom-cavity coupling may drive the
particle (or hole) to move freely on the whole lattice, which may
destroy the staggered solid order and drive the Solid-1/2 phase
directly into the SR phase. Another possibility is that the SRS
phase may be unstable and separates into domains \cite{wess}.  To
settle these issues and demonstrate explicitly the  phase diagram,
one needs an unbiased numerical simulation to go beyond mean-field
theory.

\emph{Quantum Monte Carlo (QMC) simulations.}--We adopt the
high-accuracy cluster Stochastic Series Expansion algorithm
\cite{sse1,w}. This algorithm is very efficient and has no
truncation error arising from the limitation of the photon number.
In order to distinguish  different phases, one needs to calculate
several observables: the  average excitation density $\rho=\langle
N_{\rm ex}\rangle/N$, the structure factor $S(q)/N=\langle
|\sum_{k=1}^N n_k e^{\emph{\textbf{i}}qr_k}|^2\rangle/N^2$,  and
the compressibility $\kappa=N\beta(\langle \rho^2\rangle-\langle
\rho \rangle^2)$. For the half filled Solid-1/2 phase, we take the
structure factor $S(Q)/N$ with staggered order $Q=\pi$  to
characterize the translational symmetry breaking.

\begin{acknowledgments}

\end{acknowledgments}

Fig. \ref{mcphase}(a) shows the zero-temperature phase diagram for
$\Delta=3$ and $C_6=1$, where the dashed lines of SCE match quite
well with the QMC calculations. Compared with the mean-field
results the SRS phase is greatly suppressed, indicating that
quantum fluctuations weaken the SRS order. Nonetheless, the SRS
phase remains stable in a small region as shown in the enlarged
region of the SRS phase in Fig.~\ref{mcphase}(c). In first sight
it appears difficult to reach such narrow regions of the SRS phase
experimentally.  However, this changes when considering the phase
diagram in the canonical ensemble in Fig.~\ref{mcphase}(b), where
the SRS phase exists in a wide region $\Delta\rho\approx0.1$ in
the excitation density.  In Fig.~\ref{mcphase}(b) we also find
that the region of the particle-excited SRS phase is smaller than
the hole-excited one. The  phase separated (PS) regions show that
both the Solid-1/2 and the SRS to SR phase transitions are first
order. This differs from the mean-field results and  may be
understood by the Ginzburg Landau theory \cite{xuefeng}: Because
the U(1) and translation symmetries are broken simultaneously, the
corresponding order parameters $S(Q)/N$ and $\langle
b_i^{\dag}a_i\rangle$ couple to each other and contribute to a
sixth-order term in the free energy, which results in the first
order phase transition. Moreover, a tricritical point appears
among the SRS, SR and Solid-1/2  phases, see
Fig.~\ref{mcphase}(c).

\begin{figure}[h]
\includegraphics[width=0.45\textwidth]{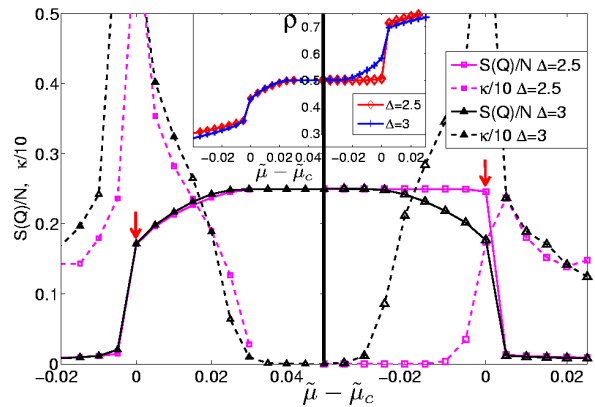}%
\caption{(color online).  The structure factor $S(Q)/N$ (solid
line) and compressibility $\kappa/10$ (dash line) along the
trajectories of the lower (left panel) and upper (right panel)
vertical cuts indicated by solid bars in Fig.~\ref{mcphase}(a).
Here, the red arrows mark the critical points with
$\tilde{\mu}=\tilde{\mu}_c$. For all cases $C_6=1$, $\beta=500$,
$N=100$, and $\Delta=3$ ($\Delta=2.5$). The inset shows the
corresponding average excitation density $\rho$. For simplicity,
the small errorbars for the observables are not shown
here.\label{ss}}
\end{figure}

To provide a convincing support for the above phase diagram, we
now concentrate on the detailed properties of the phase
transitions. Figure \ref{ss} shows  $S(Q)/N$ and  $\kappa/10$
along the trajectories  which are indicated by solid vertical
lines (red) in Fig.~\ref{mcphase}(a). Here, we take $T=C_6/500$,
which is low enough to avoid the thermal effect. In both panels,
there exist regions where a vanishing structure factor $S(Q)/N\to
0$ is accompanied by a finite $\kappa$, which means that the
translational symmetry is not broken and they are compressible,
which is characteristic of the SR phases. Regions of zero $\kappa$
with finite $S(Q)/N$ correspond to the incompressible Solid-1/2
phases. Most significantly, there exist intermediate SRS phases
which have both finite $S(Q)/N$ and $\kappa$. The corresponding
structure factor $S(Q)/N$ and average density excitation  $\rho$
have a finite jump and $\kappa$ diverges on the critical points
$\tilde{\mu}=\tilde{\mu}_c$. It signifies that the SR to SRS phase
transition is first order. In Fig.~\ref{ss}, we also show the
results for different $\Delta$. We find that, while the region of
the SRS phase in the left panel is hardly affected by $\Delta$,
the one in the right panel shrinks faster by decreasing $\Delta$.
This can be understood by the second-order atom-light interaction
process which differs for a particle or hole excited state with
corresponding second-order energies $E_p^{(2)}\sim
-\frac{g^2}{(\Delta-2C_6)}$ and
$E_h^{(2)}\sim-\frac{g^2}{\Delta}$, respectively. Note that
$|E_p^{(2)}|>|E_h^{(2)}|$, which means the particle excited SRS
phase is harder to form than the hole excited one and more easily
suppressed by decreasing $\Delta$.

\begin{figure}[h]
\includegraphics[width=0.45\textwidth]{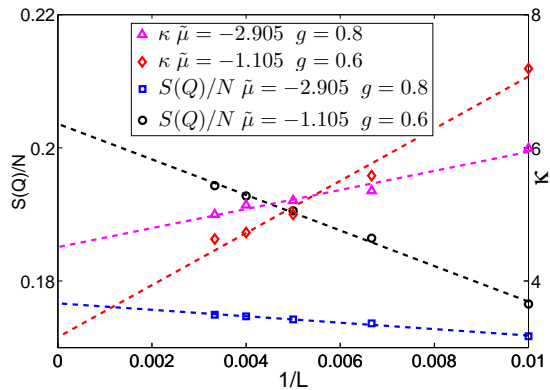}%
\caption{(color online).  The finite size scaling for  $S(Q)/N$
and $\kappa$ at $C_6=1$, $\Delta=3$ and $\beta=500$ vs different
sizes. $\tilde{\mu}=-2.905$ $(-1.105)$ correspond to the SRS
phases on the trajectories  of the lower (upper) cuts shown in
Fig.~\ref{mcphase}(a). \label{ffs}}
\end{figure}

We have also made a finite size scaling analysis for points in the
SRS phase to demonstrate that the existence of the SRS phase is
robust in the thermodynamic limit. Figure~\ref{ffs} shows the
corresponding scaling results in the SRS phase. We see that both
$S(Q)/N$ and $\kappa$ converge to finite values in the infinite
size limit.

{\it Discussion.}--Finally, we discuss two issues about the
experimental implementation. First, this system could potentially
be realized by trapping a lattice of $^{87}$Rb atoms coupled to an
ultrahigh finesse cavity \cite{Guerlin}. The $5S_{1/2}$ ground
atomic state can be coupled to the $50D_{5/2}$ Rydberg excitation
state \cite{Anderson} through a two-photon process via the
$5P_{3/2}$ intermediate state, which can be adiabatically
eliminated. The resulting effective coupling constant $g_{\rm
eff}$ is of order 10 MHZ, which is higher than both the decay
rates of the cavity and the Rydberg state. Secondly, we have
assumed in the model that only a single Rydberg excitation is
possible per site. Actually, this does not imply only one atom per
site, but instead one may work with the collective degree of
freedom of multiple atoms on each site, where only a single
Rydberg state can be excited and shared among all the atoms
\cite{Weimer2}.

{\it Conclusion.}--We have shown that the  generalized Dicke model
of a cavity QED coupled with a  Rydberg lattice gas displays a
rich phase diagram. The competition between the non-local
atom-light coupling and the atomic interaction can stabilize a
novel SRS phase. By implementing an unbiased QMC calculation, we
find that both the Solid-1/2 and SRS to SR phase transitions are
first order, and the hole-excited SRS phase is more stable than
the particle-excited one. This system may act as a new quantum
simulator for the future study of quantum many-body physics.

\begin{acknowledgments}

We acknowledge Tao Wang for helpful discussions. This work is
supported by NCET, NSFC under grants No. 11704175, 10904096, NSFB
under grants No.1092009, NKBRSFC under grants No. 2011CB921502 and
by the DFG via the SFB/Transregio 49.

\end{acknowledgments}

\bibliographystyle{apsrev}

\begin{thebibliography}{10}
\bibitem{note1} Details of the SCE calculation will be shown elsewhere.
\end{thebibliography}

\end{document}